\documentclass[pra,twocolumn,showpacs,amsmath,amssymb]{revtex4-1}
\usepackage{graphicx}
\usepackage{bm}
\usepackage{amsmath,amssymb}

\begin{document}

\title{Low-temperature coherence properties 
of $Z_2$ quantum memory}

\author{Tomoyuki Morimae}
\email{morimae@asone.c.u-tokyo.ac.jp}
\affiliation{
Department of Basic Science, University of Tokyo, 
3-8-1 Komaba, Tokyo 153-8902, Japan}
\affiliation{
Laboratoire Paul Painlev\'e, Universit\'e Lille 1,
59655 Villeneuve d'Ascq C\'edex, France
}
\date{\today}
            
\begin{abstract}
We investigate low-temperature coherence properties of the
$Z_2$ quantum memory which is
capable of storing the information of a single logical qubit.
We show
that 
the memory has superposition of macroscopically
distinct states 
for some values of a control parameter
and at sufficiently low temperature,
and that
the code states of this memory have no instability 
except for the inevitable one.
However, we also see that
the coherence power of 
this memory is limited by space and time.
We also briefly discuss the RVB memory, which is an
improvement of the $Z_2$ quantum memory,
and the relations of our results to the obscured
symmetry breaking in statistical
physics. 
\end{abstract}
\maketitle  
\section{Introduction}

Quantum memory~\cite{Kitaev,Dennis}, 
which stores the coherent information of logical qubits,
is an essential ingredient of quantum information processings,
and plays crucial roles in almost all
fields of quantum information science~\cite{Nielsen}. 
Basically, a quantum memory
consists of macroscopically many physical qubits which
encode the state of few logical qubits.
Each logical basis
should be encoded on macroscopically distinct physical states 
since otherwise the indistinguishability of 
logical bases is easily destroyed 
by local errors.
On the other hand, 
in order to store quantum coherent superposition
of these logical bases,
the memory must be able to have 
superposition of macroscopically distinct states
of physical qubits.
To maintain such superposition is very difficult.
First, if the size of the memory is infinite, 
such macroscopic superposition is impossible
since superposition of different phases is
just a classical mixture of them in 
an infinite system~\cite{Haag,SM,Alicki}.
Second, even in finite systems,
such macroscopic superposition
is usually very unstable~\cite{SM}.
Therefore, how to balance those two contradictory
demands (i.e., classical information must be encoded on 
macroscopically
distinct states but we also need superposition of them) 
is one of the most challenging problem in the implementation of
quantum memory.

The code space of a quantum memory
is often realized as the lowest-energy eigenspace
of a many-body Hamiltonian.
One of the most beautiful examples
is Kitaev's toric code~\cite{Kitaev}. 
Kitaev introduced 
a four-body Hamiltonian  
which exhibits a topological phase transition.
The ground states are
degenerated and energetically isolated from the excited states. 
Each of these ground states corresponds to different
topological phases in the thermodynamic limit, 
and therefore logical bases encoded on
these degenerate ground states are immune to local errors.
Logical qubit operations are performed by 
applying long strings of Pauli operators on physical
qubits, which means that logical operations are non-local. 

Although Kitaev's toric code is highly sophisticated
and indeed has inspired plenty of successive 
studies~\cite{Dennis,
Alicki,Alicki2,Alicki3,Takeda,Raginsky,Arakawa,Shi,Bravyi},
it is not easy
to implement a scalable toric code in a laboratory
since non-local operations are required.
A complementary approach is therefore also important from
the practical point of view.

In this paper,
we study the less elaborate but more 
feasible quantum memory, namely the $Z_2$ quantum memory.
Although the $Z_2$ quantum memory is too primitive
to be a complete and universal quantum memory,
it is still valuable to study it, since the simple structure of this memory
means feasibility in a laboratory and the possibility
of capturing the essence of theoretical aspects
of quantum memory.
In general,
a quantum memory must have
a large coherence for some values of a control parameter
in order to store a coherent information of 
logical qubits. 
Therefore,
we analyze the coherent properties of the $Z_2$
quantum memory at low temperature, 
by using the method of detecting
superposition of macroscopically distinct states
developed in Refs.~\cite{SM,magnon,visual,CP}.
We show that 
the $Z_2$ quantum memory can have superposition of 
macroscopically distinct states 
for some values of a control parameter
and at sufficiently low temperature,
and that
the code states have no instability
except for the inevitable one.
However,
we also see that
the power of superposition of macroscopically distinct states
in this memory is limited by space and time.
These results suggest
that the $Z_2$ quantum memory 
is of limited use as a prototype of a small quantum memory.

This paper is organized as follows.
In the next section, we 
briefly review the method of detecting superposition
of macroscopically distinct states.
In Sec.~\ref{zero}, we study the zero-temperature case.
We next study the finite-temperature case
in Sec.~\ref{finite}.
Finally, 
in Sec.~\ref{discussionRVB},
we briefly discuss the RVB quantum memory,
which is an improvement of the $Z_2$ quantum memory,
and the relations of our results
to the concept of symmetry breaking in statistical physics
in Sec.~\ref{conclusion}.

\section{Index $p$ and VCM}

In this section, we briefly
review the method of detecting superposition of macroscopically
distinct states in quantum many-body states~\cite{SM,magnon,visual,CP}. 

Let us consider an $N$-site lattice
($1\ll N<\infty$) where
the dimension of the Hilbert space on each site is 
an $N$-independent constant,
such as a chain of $N$ spin-1/2 particles.
Throughout this paper,
$f(N)=O(N^k)$ means
\begin{eqnarray*}
\lim_{N\to\infty}\frac{f(N)}{N^k}=\mbox{const.}\neq0.
\end{eqnarray*}

For a given pure state $|\psi\rangle$, 
the index $p$ $(1\leq p \leq 2)$ is defined by
\begin{eqnarray*}
\max_{\hat{A}}
[\langle\psi|\hat{A}^2|\psi\rangle
-\langle\psi|\hat{A}|\psi\rangle^2]
=O(N^p),
\end{eqnarray*}
where the maximum 
is taken over all Hermitian additive operators 
$\hat{A}$.
Here, an additive operator 
\begin{eqnarray*}
\hat{A}
=\sum_{l=1}^N\hat{a}(l)
\end{eqnarray*}
is a sum of local operators $\{\hat{a}(l)\}_{l=1}^N$,
where $\hat{a}(l)$
is a local operator acting on site $l$.
For example, if the system is a chain 
of $N$ spin-$1/2$ particles,
$\hat{a}(l)$ is a linear combination
of three Pauli operators, 
$\hat{\sigma}_x(l),\hat{\sigma}_y(l),\hat{\sigma}_z(l)$, 
and the identity operator $\hat{1}(l)$ acting on site $l$.
In this case, the $x$-component of the total
magnetization 
\begin{eqnarray*}
\hat{M}_x\equiv\sum_{l=1}^N\hat{\sigma}_x(l)
\end{eqnarray*}
and the $z$-component of the total staggard magnetization
\begin{eqnarray*}
\hat{M}_z^{st}\equiv\sum_{l=1}^N(-1)^l\hat{\sigma}_z(l)
\end{eqnarray*}
are, for example, additive operators.
The index $p$ takes the minimum value 1 for any 
product state
\begin{eqnarray*}
\bigotimes_{l=1}^N|\phi_l\rangle,
\end{eqnarray*}
where $|\phi_l\rangle$ is a state of site $l$
(this means that $p>1$ is an entanglement witness
for pure states). 
If $p$ takes the maximum value 2,
the state contains superposition of macroscopically 
distinct states
because in this case 
the relative fluctuation of an additive operator
does not vanish in
the thermodynamic limit:
\begin{eqnarray*}
\lim_{N\to\infty}
\frac{\sqrt{\langle\psi|\hat{A}^2|\psi\rangle
-\langle\psi|\hat{A}|\psi\rangle^2}}{N}
\neq0,
\end{eqnarray*}
and because the fluctuation of an observable
in a pure state means the existence of a superposition
of eigenvectors of that observable
corresponding to different eigenvalues.

For example, the $N$-qubit GHZ state 
\begin{eqnarray*}
|\mbox{GHZ}\rangle\equiv
\frac{1}{\sqrt{2}}(|0^{\otimes N}\rangle+|1^{\otimes N}\rangle), 
\end{eqnarray*}
which obviously contains
superposition of macroscopically distinct states,
has $p=2$, since
\begin{eqnarray*}
\langle\mbox{GHZ}|\hat{M}_z^2|\mbox{GHZ}\rangle
-\langle\mbox{GHZ}|\hat{M}_z|\mbox{GHZ}\rangle^2
=O(N^2).
\end{eqnarray*}

It was shown in Ref.~\cite{SM} that a state having $p=2$
is unstable against a local noise from the environment and
a local measurement, whereas a state having $p=1$ is stable.


There is an efficient method of calculating
index $p$~\cite{magnon,visual}.
For simplicity, we assume that 
the Hilbert space on each site is two-dimensional one.
Generalizations to higher dimensional cases are immediate.

For a given pure state $|\psi\rangle$,
let us define the $3N\times3N$ Hermitian matrix called 
the variance-covariance matrix (VCM) by
\begin{eqnarray*}
V_{\alpha l, \beta l'}\equiv
\langle\psi|\hat{\sigma}_{\alpha}(l)\hat{\sigma}_{\beta}(l')|\psi\rangle
-\langle\psi|\hat{\sigma}_{\alpha}(l)|\psi\rangle
\langle\psi|\hat{\sigma}_{\beta}(l')|\psi\rangle,
\end{eqnarray*}
where 
$\alpha,\beta=x,y,z$;
$l,l'=1,2,\cdots,N$; 
$\hat{\sigma}_x(l)$, $\hat{\sigma}_y(l)$, and $\hat{\sigma}_z(l)$
are Pauli operators on site $l$.
Since the VCM is Hermitian, all eigenvalues are real.
Let $e_1$ be the largest eigenvalue 
of the VCM. Then 
\begin{eqnarray*}
e_1 = O(N^{p-1})
\end{eqnarray*}
is satisfied~\cite{magnon,visual},
which means
that we have only to calculate $e_1$ to obtain the value
of index $p$.
Since a matrix of a polynomial size can be diagonalized 
within a polynomial steps,
$e_1$ is obtained efficiently by numerical calculations.

\section{Zero temperature}
\label{zero}
Let us first study the zero-temperature case.
We consider the one-dimensional periodic chain
of $N$ qubits.
The code space of the $Z_2$ quantum memory is stabilized by
the ``bond operators" $(l=1,2,...,N)$~\cite{Alicki3}
\begin{eqnarray*}
\hat{B}_l\equiv\hat{\sigma}_z(l)\hat{\sigma}_z(l+1),
\end{eqnarray*}
which act on the ``virtual qubits" (or ``dual qubits") embedded on bonds,
where $\hat{\sigma}_z(N+1)=\hat{\sigma}_z(1)$.
Since
\begin{eqnarray*}
\hat{B}_N=\prod_{l=1}^{N-1}\hat{B}_l,
\end{eqnarray*}
stabilizers of the $Z_2$ quantum memory are generated by $N-1$ bond
operators $\{\hat{B}_1,\hat{B}_2,...,\hat{B}_{N-1}\}$, 
which means that the code space is 
$2^{N-(N-1)}=2^1$ dimensional subspace.
The centralizers of these stabilizers are generated by
\begin{eqnarray}
\hat{X}\equiv\prod_{l=1}^N\hat{\sigma}_x(l),\label{logicalX}
\end{eqnarray}
and
\begin{eqnarray*}
\hat{Z}&\equiv&\hat{\sigma}_z(1).
\end{eqnarray*}
They work as the logical bit flip and logical phase, respectively.
The code space is also specified as the 
lowest-energy eigenspace of the two-body Hamiltonian
\begin{eqnarray*}
\hat{H}_0=-\sum_{l=1}^N\hat{B}_l.
\end{eqnarray*}
It is obvious that two degenerate separable ground states
of this Hamiltonian are macroscopically distinct
with each other, and therefore the indistinguishability of logical bases,
$|\tilde{0}\rangle$ and $|\tilde{1}\rangle$,
is not destroyed by local errors.

Ideally, the logical Hadamard operation 
\begin{eqnarray*}
|\tilde{0}\rangle&\to&
\frac{1}{\sqrt{2}}\Big(|\tilde{0}\rangle+|\tilde{1}\rangle\Big)\\
|\tilde{1}\rangle&\to&
\frac{1}{\sqrt{2}}\Big(|\tilde{0}\rangle-|\tilde{1}\rangle\Big),
\end{eqnarray*}
which is the essential ingredient of various quantum information
processings,
is realized by using the logical $\hat{X}$ operation,
Eq.~(\ref{logicalX}).
However, such non-local operation is
not easy to experimentally implement.
Therefore,
it is reasonable to try to manipulate
the quantum memory in the local way:
\begin{eqnarray}
\hat{H}=\hat{H}_0
+\lambda\sum_{l=1}^N\hat{\sigma}_x(l),
\label{Hoftim}
\end{eqnarray}
where $\lambda$ is an external control parameter.

\begin{figure}[htbp]
\begin{center}
\includegraphics[width=0.5\textwidth]{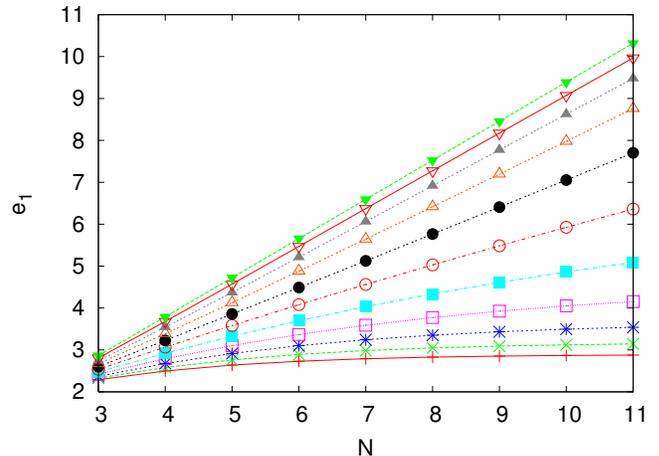}
\end{center}
\caption{(Color online) 
$e_1$ versus $N$ for $|E_0(\lambda)\rangle$ 
with various $\lambda$. From the bottom,
$\lambda=$1.5, 1.4, 1.3, 1.2, 1.1, 1.0, 0.9, 0.8, 0.7, 0.6, and 0.5, respectively. Lines are guides to the eye.}
\label{emax.eps}
\end{figure}

From the Perron-Frobenius theorem~\cite{Horn}, the ground state
of this Hamiltonian is non-degenerate if $\lambda\neq0$.
It is also known that
this Hamiltonian
exhibits the quantum phase transition
at $\lambda=1$~\cite{Sachdev}.
Let us denote the exact ground
state of $\hat{H}$ corresponding to the external parameter $\lambda$
by $|E_0(\lambda)\rangle$,
and evaluate index $p$ of $|E_0(\lambda)\rangle$
for various $\lambda$.
In Fig.~\ref{emax.eps},
the largest eigenvalue 
$e_1$ of the VCM versus $N$ is plotted by changing
the value of $\lambda$.
This figure shows that 
\begin{itemize}
\item
$|E_0(\lambda\ge1)\rangle$ has $p<2$
\item
$|E_0(\lambda<1)\rangle$ has $p=2$,
\end{itemize}
which means that
the $Z_2$ quantum memory has superposition
of macroscopically distinct states for 
values $\lambda<1$ of the control parameter $\lambda$.

\begin{figure}[htbp]
\begin{center}
\includegraphics[width=0.5\textwidth]{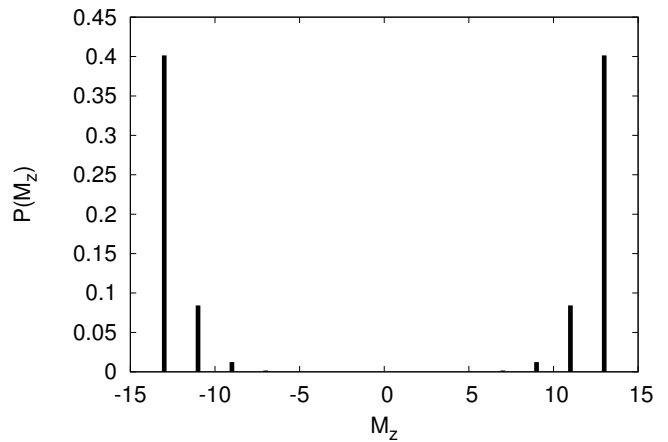}
\end{center}
\caption{The probability distribution $P(M_z)$ of $\hat{M}_z$ in
$|E_0(0.5)\rangle$ with $N=13$.
Similar structures are obtained for other values of $N$.
}
\label{13dataP_z.eps}
\end{figure}

In order to see the structure of the superposition,
the probability distribution $P(M_z)$ of $\hat{M}_z$ in
$|E_0(0.5)\rangle$ 
is plotted
in Fig.~\ref{13dataP_z.eps}
for $N=13$.
This figure suggests that
the ground state
is, in a rough picture,
a superposition of two macroscopically distinct states:
\begin{eqnarray}
|E_0(0.5)\rangle\simeq|\phi_+\rangle+|\phi_-\rangle,
\label{E_0}
\end{eqnarray}
where
$|\phi_\pm\rangle$ are some states satisfying
$\langle\phi_\pm|\hat{M}_z|\phi_\pm\rangle\simeq\pm N$, 
respectively.

\begin{figure}[htbp]
\begin{center}
\includegraphics[width=0.5\textwidth]{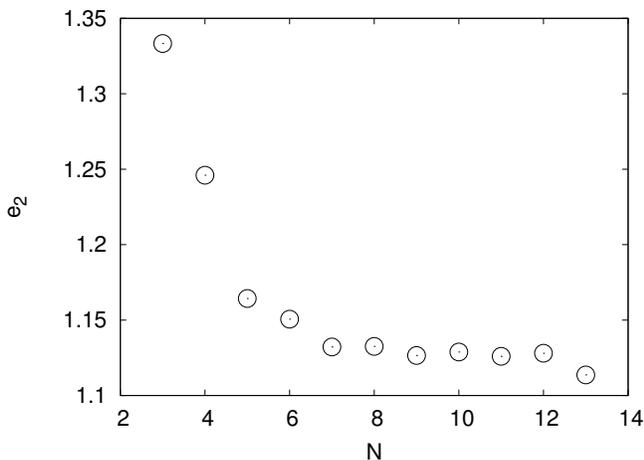}
\end{center}
\caption{$e_2$ versus $N$ for $|E_0(0.5)\rangle$.}
\label{tim_e2.eps}
\end{figure}

By seeing the eigenvector of the VCM 
corresponding to the largest eigenvalue $e_1$,
we can also know that the additive operator which
gives the maximum fluctuation in $|E_0(\lambda<1)\rangle$ is 
\begin{eqnarray*}
\hat{M}_z\equiv\sum_{l=1}^N\hat{\sigma}_z(l).
\end{eqnarray*}
According to Ref.~\cite{SM}, 
this means that
$|E_0(\lambda<1)\rangle$ 
is unstable against the local noise
described by the interaction Hamiltonian
\begin{eqnarray}
\hat{H}_{\rm int}\equiv
\sum_{l=1}^Nf(l)\hat{\sigma}_z(l),
\label{unstablenoise}
\end{eqnarray}
where $f(l)$ is a noise parameter of a long wavelength.
Although
this noise
is inevitable since we need the superposition
of macroscopically distinct logical bases,
we can still show that 
$|E_0(\lambda<1)\rangle$ 
has no other instability than this inevitable one.
In Fig.~\ref{tim_e2.eps}, we plot the second largest eigenvalue $e_2$ 
of the VCM versus $N$
for $|E_0(0.5)\rangle$. From this
figure we obtain 
\begin{eqnarray*}
e_2\le O(N^0),
\end{eqnarray*}
which means that
$|E_0(0.5)\rangle$ is 
stable against all noises
of the type
\begin{eqnarray*}
\hat{H}_{\rm int}\equiv\sum_{l=1}^Nf(l)\hat{a}(l)
\end{eqnarray*}
except for the case of Eq.~(\ref{unstablenoise}).

In short, we have seen that the $Z_2$ quantum memory
can have superposition of macroscopically distinct states
and it is stable against any local noise
except for the inevitable one.

\begin{figure}[htbp]
\begin{center}
\includegraphics[width=0.5\textwidth]{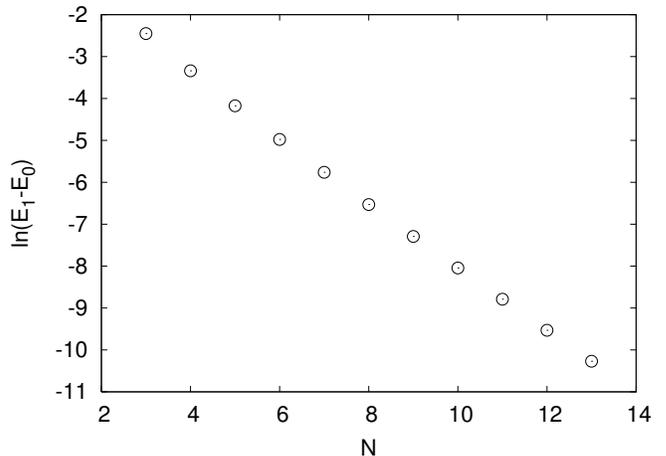}
\end{center}
\caption{$\ln\Delta E$ versus $N$ for $\lambda=0.5$.
}
\label{tim_gap.eps}
\end{figure}

As is seen in Fig.~\ref{tim_gap.eps}, 
the energy gap $\Delta E\equiv E_1-E_0$ 
between the exact ground state and the first excited state
for $\lambda=0.5$ decays exponentially fast as $N\to\infty$.
Therefore, it is physically allowed to take
a linear combination of the exact ground state
and the first excited state for sufficiently
large $N$.

From Eq.~(\ref{E_0}) and the analogy with the 
physics of the single two-level atom system,
the rough picture of the
first excited state $|E_1(0.5)\rangle$ is expected
to be
\begin{eqnarray}
|E_1(0.5)\rangle\simeq
|\phi_+\rangle-|\phi_-\rangle,
\label{E_1}
\end{eqnarray}
and therefore the superposition
\begin{eqnarray*}
|E_0'\rangle\equiv
\frac{1}{\sqrt{2}}\Big(
|E_0(0.5)\rangle+|E_1(0.5)\rangle\Big)
\end{eqnarray*}
of the exact ground state and the first excited state
is expected to have no
superposition of macroscopically distinct
states.

\begin{figure}[htbp]
\begin{center}
\includegraphics[width=0.5\textwidth]{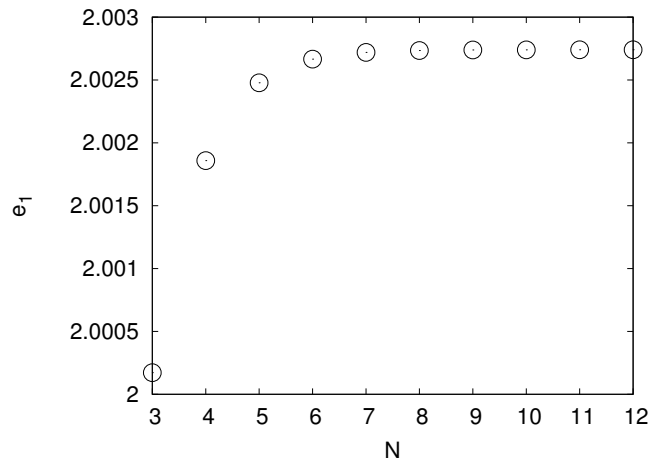}
\end{center}
\caption{$e_1$ versus $N$ for $|E_0'\rangle$}
\label{tim_1data_2data_emax.eps}
\end{figure}

Indeed, we plot $e_1$ 
versus $N$ for $|E_0'\rangle$
in Fig.~\ref{tim_1data_2data_emax.eps}.
This figure shows that $e_1=O(N^0)$, which means that
the superposition of macroscopically distinct states
in quantum memory disappears for sufficiently large
system size.
Therefore, the $Z_2$ quantum memory is of limited use
as a small quantum memory.

It is worth mentioning that
such a fast decay of the energy gap is a natural consequence
of the locality of the Hamiltonian.
According to Eqs.~(\ref{E_0}) and (\ref{E_1}), the difference between
the exact ground state and the first excited state
is the relative phase of $|\phi_\pm\rangle$.
If the energy gap does not go to 0 for large $N$, we can distinguish
the exact ground state and the first excited state
by measuring the energy.
If the Hamiltonian is a local one,
i.e., it does not contain many-point correlation,
this means that we can know the relative phase
by measuring only few-point correlations, which obviously contradicts
to the common sense of decoherence~\cite{Wakita}.

In addition to the limit of the space, the exponential decay
of the energy gap also gives the limit to the operation time.
Assume that we perform the logical Hadamard gate
by adiabatically increasing the control parameter $\lambda$.
Then, according to the theory of adiabatic 
quantum computation~\cite{Farhi,Childs},
the exponential decay of the energy gap means
the exponentially long operation time:
\begin{eqnarray*}
T\simeq\frac{1}{(\Delta E)^2},
\end{eqnarray*}
where $\Delta E$ is the minimum energy gap.
Therefore, the large $Z_2$ quantum memory also has the limit of the
operation time.
\section{Finite temperature}
\label{finite}

Next let us study
the $Z_2$ quantum memory at finite temperature.

At finite temperature $T$, a system is generally in
the equilibrium state: 
\begin{eqnarray*}
\hat{\rho}=\frac{e^{-\hat{H}/kT}}{\mbox{Tr}(e^{-\hat{H}/kT})}.
\end{eqnarray*}
If the state is not necessarily pure, 
index $p$ cannot detect superposition
of macroscopically distinct states
since a fluctuation is not necessarily equivalent to
the coherence in mixed states. 

In order to detect superposition of macroscopically distinct
states in mixed states, index $q$ was proposed 
in Ref.~\cite{SM05}.
For a given many-body state $\hat{\rho}$,
index $q$ ($1\le q\le 2$) is defined by
\begin{eqnarray*}
\max\Big(N,\max_{\hat{A}}
\Big\|[\hat{A},[\hat{A},\hat{\rho}]]\Big\|_1\Big)
=O(N^q),
\end{eqnarray*}
where $\|\hat{X}\|_1\equiv
\mbox{Tr}\sqrt{\hat{X}^\dagger\hat{X}}$ is the 1-norm, and 
$\max_{\hat{A}}$ means the maximum over all
Hermitian additive operators $\hat{A}$.
As detailed in Ref.~\cite{SM05}, 
$q$ takes the minimum value 1 for any separable state,
\begin{eqnarray*}
\sum_i\lambda_i\bigotimes_{l=1}^N
|\phi_l^{(i)}\rangle
\langle\phi_l^{(i)}|,
\end{eqnarray*}
where $|\phi_l^{(i)}\rangle$ is a state of site $l$.
On the other hand, if $q$ takes the maximum value 2,
the state contains superposition
of macroscopically distinct states.
In particular,
for pure states, $p=2\iff q=2$.

Unlike the case of index $p$,
there is no
method of efficiently calculating index $q$ at the time of writing.
However, we can calculate a lower bound of the value of
$q$.
Indeed, let us note that
\begin{eqnarray*}
\Big\|\big[\hat{A},\hat{\rho}\big]\Big\|_2^2&=&
\mbox{Tr}\Big(\big[\hat{A},\hat{\rho}\big]^\dagger\big[\hat{A},\hat{\rho}\big]\Big)\\
&=&\mbox{Tr}\Big(\hat{\rho}\big[\hat{A},\big[\hat{A},\hat{\rho}\big]\big]\Big)\\
&\le&\frac{\mbox{Tr}(\hat{\rho}[\hat{A},[\hat{A},\hat{\rho}]])}
{\|\hat{\rho}\|_\infty}\\
&\le&\Big\|\big[\hat{A},
\big[\hat{A},\hat{\rho}\big]\big]\Big\|_1,
\end{eqnarray*}
where 
$\|\hat{X}\|_2\equiv\sqrt{\mbox{Tr}(\hat{X}^\dagger\hat{X})}$ 
is the 2-norm,
$\|\hat{X}\|_\infty$ is the operator norm.
If we define the
$3N\times3N$ Hermitian matrix $W$ by
\begin{eqnarray}
W_{\alpha,l,\beta,l'}\equiv
\mbox{Tr}\Big(\big[\hat{\rho},\hat{\sigma}_\alpha(l)\big]
\big[\hat{\sigma}_\beta(l'),\hat{\rho}\big]\Big),
\label{q_2'vcm}
\end{eqnarray}
where $\alpha,\beta=x,y,z$ and $l,l'=1,2,...,N$,
it is easy to see that  
the order of
\begin{eqnarray*}
\max_{\hat{A}}\Big\|[\hat{A},\hat{\rho}]\Big\|_2^2 
\end{eqnarray*}
with respect to $N$
is equal to that of $e_1N$, where $e_1$ is the largest eigenvalue of $W$.

\begin{figure}[htbp]
\begin{center}
\includegraphics[width=0.5\textwidth]{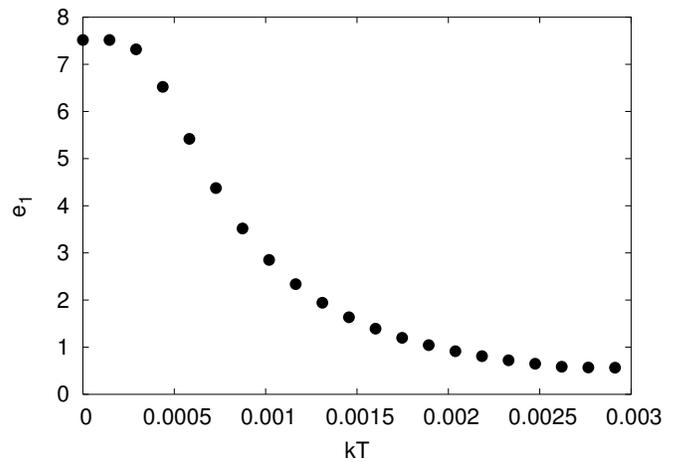}
\end{center}
\caption{$e_1$ versus $kT/J$ for $N=8$ ($J=1$).}
\label{timfig.eps}
\end{figure}

In Fig.~\ref{timfig.eps}, we plot
the largest eigenvalue $e_1$ of the matrix 
$W$, Eq.~(\ref{q_2'vcm}),
versus $kT$ for the equilibrium state
of the Hamiltonian Eq.~(\ref{Hoftim}),
for $\lambda=0.5$ and $N=8$.
This figure shows that superposition of macroscopically
distinct states at zero temperature
persists at sufficiently low temperature.

\section{RVB memory}
\label{discussionRVB}

In this paper, we have seen that 
the $Z_2$ quantum memory can have superposition
of macroscopically distinct states for 
values $\lambda<1$ of the control parameter $\lambda$,
and this superposition of macroscopically distinct
states is stable against any local noise except for
the inevitable one.
Let us briefly discuss a possibility of avoiding this inevitable
instability without introducing too much sophisticated
methods. 

\begin{figure}[htbp]
\begin{center}
\includegraphics[width=0.5\textwidth]{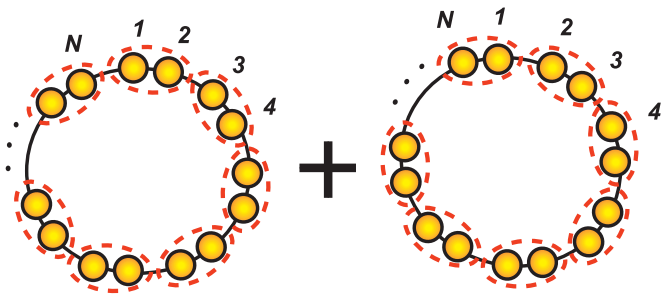}
\end{center}
\caption{(Color online) The nearest-neighbor RVB state on a one-dimensional periodic
lattice. 
Red circles represent singlet pairs.
Two macroscopically distinct VB states are superposed.}
\label{rvb}
\end{figure}

One of the most simple ways would be
to use RVB (Resonating Valence Bond) states:
\begin{eqnarray*}
|\Psi\rangle&\equiv&
\frac{1}{\sqrt{2+4(-\frac{1}{2})^{N/2}}}
\Big[
\bigotimes_{l=1}^{N/2}|2l-1,2l\rangle
+\bigotimes_{l=1}^{N/2}|2l,2l+1\rangle
\Big]\\
&\simeq&
\frac{1}{\sqrt{2}}
\bigotimes_{l=1}^{N/2}|2l-1,2l\rangle
+\frac{1}{\sqrt{2}}
\bigotimes_{l=1}^{N/2}|2l,2l+1\rangle\\
&\equiv&\frac{1}{\sqrt{2}}|{\rm VB}_1\rangle+
\frac{1}{\sqrt{2}}|{\rm VB}_2\rangle,
\end{eqnarray*}
where $|i,j\rangle$ represents the singlet pair between
sites $i$ and $j$, and the periodic boundary condition
is assumed (Fig.~\ref{rvb}).
This state is realized as a ground state of the one-dimensional
spin ladder model or the Majumdar-Ghosh model~\cite{MG}.

$|\Psi\rangle$ is a 
superposition of two macroscopically distinct
symmetry-broken states,
$|{\rm VB}_1\rangle$ and $|{\rm VB}_2\rangle$.
The advantage of this state is that,
as is shown in Appendix,
there is no long-range two-point correlation in this state:
\begin{eqnarray}
\langle \Psi|\hat{a}(l)\hat{a}(l')|\Psi\rangle
-\langle \Psi|\hat{a}(l)|\Psi\rangle
\langle \Psi|\hat{a}(l')|\Psi\rangle=0,
\label{nolongrange}
\end{eqnarray}
for any pair of local operators $\hat{a}(l)$ and 
$\hat{a}(l')$ with $|l-l'|\ge 2$.
Therefore, $|\Psi\rangle$ has $p=1$ and this means
that the state is stable against any local noise of the type
\begin{eqnarray*}
\hat{H}_{\rm int}=\sum_{l=1}^Nf(l)\hat{a}(l).
\end{eqnarray*}

In order to detect superposition of macroscopically
distinct states in $|\Psi\rangle$,
we must consider the sum of bilocal operators. 
Indeed, let us consider the operator
\begin{eqnarray*}
\hat{T}\equiv\sum_{l=1}^N(-1)^l\hat{t}_{l,l+1},
\end{eqnarray*}
where
\begin{eqnarray*}
\hat{t}_{l,l+1}\equiv
|l,l+1\rangle\langle l,l+1|
\end{eqnarray*}
is the projection operator on the singlet state of
sites $l$ and $l+1$.
Then,
we can show that the state $|\Psi\rangle$ 
has superposition of macroscopically
distinct states in the sense of
\begin{eqnarray}
\langle \Psi|\hat{T}^2|\Psi\rangle
-\langle \Psi|\hat{T}|\Psi\rangle^2
=O(N^2).
\label{rfluc}
\end{eqnarray}
A proof is given in Appendix.

In summary, we have seen that the stability
of the $Z_2$ quantum memory can be improved by introducing local entanglement
between nearest neighbor sites.
The RVB state thus created still has superposition of macroscopically
distinct states.

\section{Conclusion and Discussion}
\label{conclusion}

In this paper, we have 
investigated the low-temperature coherence properties of the
$Z_2$ quantum memory.
We have shown
that 
(i) the memory can have superposition of macroscopically
distinct states 
for the values $\lambda<1$ of the control parameter $\lambda$
and at sufficiently low temperature,
(ii) the code states of this memory have no instability 
except for the inevitable one,
and
(iii) the power of superposition of macroscopically distinct states
in this memory is limited by space and time.
We have also briefly discussed how the RVB memory
improves the $Z_2$ quantum memory.

To conclude this paper, 
let us briefly discuss the relations of our results
to symmetry breaking in statistical physics.

Symmetry breaking is one of the most fundamental concepts
in modern physics~\cite{Nambu,Weiss,Duerr}.
According to Landau-Ginzburg theory~\cite{Landau}, 
the state of the phase is a local minimum
of the effective potential as a function of the order parameter:
if the temperature is high, the potential
has the unique minimum at the origin
and therefore the system is in the symmetric phase,  
whereas
at sufficiently low temperature, the effective potential 
becomes the double-well type whose
local minima correspond to the symmetry-broken phases.

Although such intuitive picture of the symmetry-breaking
has contributed to the progress of physics
for long time, 
it has been often pointed out
by many researchers that
such picture is not always 
correct if the system is of a 
finite volume~\cite{Koma,Horsch,Oitmaa,Momoi,SMU(1),Mukaida}.
For example, if the many-body Hamiltonian $\hat{H}$
and the order operator $\hat{O}$ do not commute with
each other $[\hat{H},\hat{O}]\neq0$
(typical examples in condensed matter physics are the transverse Ising model,
the Heisenberg antiferromagnet,
and the Hubbard model),
the exact ground state of a finite volume is often non-degenerate
and therefore symmetric.
This symmetric exact ground state is completely different
to the symmetry-broken ``mean-field ground states",
which are inherently separable
since the mean-field approximation neglects
the correlations among sites~\cite{Nakajima}.

Such symmetric exact ground state $|E_0\rangle$
often
has the peculiar property 
that the relative fluctuation of a macroscopic observable $\hat{A}$,
which is mostly the order operator $\hat{O}$,
does not vanish even in the
thermodynamic limit:
\begin{eqnarray*}
\lim_{N\to\infty}
\frac{\sqrt{\langle E_0|\hat{A}^2|E_0\rangle
-\langle E_0|\hat{A}|E_0\rangle^2}}{N}
\neq0,
\end{eqnarray*}
where $N$ is the number of total sites (alias 
the volume of the system).
Since this means that a macroscopic observable
does not have a definite value even in the thermodynamic limit,
$|E_0\rangle$ is an anomalous state~\cite{SM} from the view point
of thermodynamics where
any macroscopic observable is supposed to
have definite value~\cite{SM,Landau}.
In terms of index $p$, $|E_0\rangle$ has $p=2$ and
therefore contains superposition of macroscopically distinct states.
In other words, the symmetry of the ground state
is ``obscured" by such large quantum fluctuation. 
This effect is often called ``obscured symmetry breaking"~\cite{Koma}.

However,
when the exact ground state has such anomalous property,
it is often the case 
that the energy gap between the exact ground state
and the low-lying eigenstates
decays very fast as $N\to\infty$~\cite{Koma,Horsch}.
Then, it is physically allowed to take a linear combination 
of the exact ground state and some of the low-lying eigenstates
to form an approximate ground state $|E_0'\rangle$.
Thus constructed $|E_0'\rangle$ are believed to
break the symmetry 
and be ``ergodic" in the sense that
\begin{eqnarray*}
\lim_{N\to\infty}
\frac{\sqrt{\langle E_0'|\hat{A}^2|E_0'\rangle
-\langle E_0'|\hat{A}|E_0'\rangle^2}}{N}
=0
\end{eqnarray*}
for any macroscopic observable $\hat{A}$~\cite{Koma,Horsch}.
In terms of index $p$,
this means that $|E_0'\rangle$ has no superposition of macroscopically
distinct states since $p<2$.

Indeed, Horsch and Linden~\cite{Horsch}
introduced the trial state $\hat{O}|E_0\rangle$
which approximates the first excited state,
and showed that the energy gap between the exact ground
state and the trial state decays as fast as or faster than $1/N$.
They also showed that a linear combination 
of the exact ground
state and the trial state 
exhibits the desired
$Z_2$ symmetry breaking. 

Koma and Tasaki~\cite{Koma} rigorously showed
that the linear combination $|\psi\rangle$ of the exact ground state
and the trial state is also the ergodic state
in the sense that
\begin{eqnarray}
\frac{1}{N^2}
\Big[
\langle \psi|\hat{A}^2|\psi\rangle
-\langle \psi|\hat{A}|\psi\rangle^2
\Big]=0
\label{Komashowed}
\end{eqnarray}
for any translationally invariant $\hat{A}$.

As is pointed out in Ref.~\cite{Raginsky},
the low-temperature coherence properties of quantum memory
is closely related to the obscured symmetry breaking
in statistical physics.
Indeed,
our results in this paper are considered as
improvement of the previous results,
since Hamiltonian Eq.~(\ref{Hoftim}) is equivalent to
that of the transverse Ising model~\cite{Chakrabarti}
in condensed matter physics.
(For example, this model was used to describe the order-disorder 
transition in some double-well ferroelectric systems, 
such as potassium dihydrogen 
phosphate ($\rm{KH}_2\rm{PO}_4$) crystals~\cite{Gennes}.)

First, by numerical calculations, we have explicitly shown
for the first time how the macroscopic coherence
properties of the exact ground state changes when the
transverse magnetic field is changed (Fig.~\ref{emax.eps}).
We have also visualized the structure of the macroscopic
superposition in the exact ground state
(Fig.~\ref{13dataP_z.eps}), and seen that the
exact ground state is approximately an equal weight 
superposition of two symmetry-broken phases.

Second, we have shown
that only $\hat{M}_z$ fluctuates macroscopically
in the $\lambda<1$ phase (Fig.~\ref{tim_e2.eps}).
Since the ground state is symmetric, 
this means that
the second moment of $\hat{M}_z$ is of $O(N^2)$
in that phase.
In terms of statistical physics, this means 
that $\hat{M}_z$ is the unique order operator in this phase,
which is, to the author's knowledge, a new result.

Third, we have shown that the equal-weight superposition
of the exact ground state and the first excited state has $p=1$ 
(Fig.~\ref{tim_1data_2data_emax.eps}).  
This means that the superposition of the exact ground state
and the first excited state is ergodic.
Although similar results have been obtained,
the advantages of our result are
(i) instead of the trial state, we have directly used the first
excited state to show the ergodicity,
(ii)
our result that the fluctuation is of $O(N)$
is stronger than Eq.~(\ref{Komashowed}),
and (iii)
we have shown the ergodicity for any
additive operator, whereas only translationally invariant
additive operators are considered in the previous studies.
(A disadvantage of our result is that
it is less general since we have used numerical 
calculations.)
In summary, to consider index $p$ for the ground states of many-body
Hamiltonians in condensed matter physics
is very useful for the study of the foundation of statistical
physics.

\acknowledgements
The author thanks A. Shimizu and Y. Matsuzaki
for useful discussions.
This work was partially supported by Japan
Society for the Promotion of Science. 

\appendix*
\section{}
\subsection{Proof of Eq.~(\ref{nolongrange})}
Note that 
$|\Psi\rangle$ is 
a simultaneous eigenvector of $\hat{M}_x$, $\hat{M}_y$, 
and $\hat{M}_z$ corresponding to the eigenvalue 0, since a singlet 
is a simultaneous eigenvector of $x$, $y$, and $z$ component
of the total magnetization corresponding to the eigenvalue 0.
Since each of the states
\begin{eqnarray*}
\hat{\sigma}_x(l)|\Psi\rangle,\\
\hat{\sigma}_y(l)|\Psi\rangle,\\
\hat{\sigma}_x(l)\hat{\sigma}_x(l')|\Psi\rangle,\\
\hat{\sigma}_x(l)\hat{\sigma}_y(l')|\Psi\rangle,\\ 
\hat{\sigma}_x(l)\hat{\sigma}_z(l')|\Psi\rangle,\\ 
\hat{\sigma}_y(l)\hat{\sigma}_y(l')|\Psi\rangle,\\ 
\hat{\sigma}_y(l)\hat{\sigma}_z(l')|\Psi\rangle, 
\end{eqnarray*}
for $|l-l'|\ge2$
has no component in the eigenspace of $\hat{M}_z$ corresponding
to the eigenvalue $M_z=0$, they are orthogonal to $\langle\Psi|$.
In the same way,
\begin{eqnarray*}
\hat{\sigma}_z(l)|\Psi\rangle,\\
\hat{\sigma}_z(l)\hat{\sigma}_z(l')|\Psi\rangle,
\end{eqnarray*}
for $|l-l'|\ge2$
are orthogonal to $\langle\Psi|$ since each of them
has no component
in the eigenspace of $\hat{M}_x$ corresponding to
the eigenvalue $M_x=0$.
Hence we have shown Eq.~(\ref{nolongrange}).~\rule[-2pt]{5pt}{10pt}

\subsection{Proof of Eq.~(\ref{rfluc})}
\begin{figure}[htbp]
\begin{center}
\includegraphics[width=0.2\textwidth]{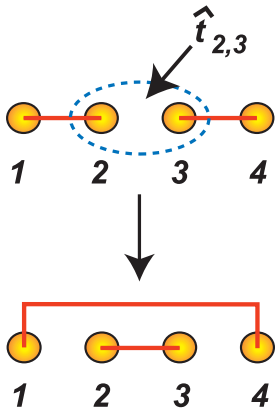}
\end{center}
\caption{(Color online) The projection operator $\hat{t}_{2,3}$ swaps the entanglement.
Red lines represent singlet pairs.
}
\label{swap}
\end{figure}

Before showing the equation, let us note some useful relations.
First, the projection operator
$\hat{t}_{2,3}$ ``swaps" the entanglement
\begin{eqnarray*}
\hat{t}_{2,3}
|\circ\circ\bullet\bullet\rangle
=-\frac{1}{2}
|\circ\bullet\bullet\circ\rangle,
\end{eqnarray*}
where singlet pairs are schematically represented:
sites represented by the circle of the
same color make a singlet pair
(see Fig.~\ref{swap}).

Second, by using this equation,
we obtain
\begin{eqnarray*}
\langle\circ\bullet\bullet\circ
|\circ\circ\bullet\bullet\rangle
&=&
\langle\circ\bullet\bullet\circ|
\hat{t}_{2,3}
|\circ\circ\bullet\bullet\rangle\\
&=&-\frac{1}{2}
\langle\circ\bullet\bullet\circ
|\circ\bullet\bullet\circ\rangle\\
&=&-\frac{1}{2}
\end{eqnarray*}
and
\begin{eqnarray*}
\langle\circ\circ\bullet\bullet|
\hat{t}_{2,3}
|\circ\circ\bullet\bullet\rangle
&=&-\frac{1}{2}
\langle\circ\circ\bullet\bullet
|\circ\bullet\bullet\circ\rangle\\
&=&\frac{1}{4}.
\end{eqnarray*}

Third, by iterating the swap process, 
\begin{eqnarray*}
|{\rm VB_2}\rangle=(-2)^{N/2-1}
\prod_{l=1}^{N/2-1}\hat{t}_{2l,2l+1}|{\rm VB_1}\rangle,
\end{eqnarray*}
and therefore
\begin{eqnarray*}
\langle{\rm VB_2}|{\rm VB_2}\rangle=(-2)^{N/2-1}
\langle{\rm VB_2}|
\prod_{l=1}^{N/2-1}\hat{t}_{2l,2l+1}|{\rm VB_1}\rangle,
\end{eqnarray*}
which gives
\begin{eqnarray*}
\langle{\rm VB_2}|{\rm VB_1}\rangle
=\Big(-\frac{1}{2}\Big)^{N/2-1}\simeq0.
\end{eqnarray*}

Let us define
\begin{eqnarray*}
|\phi_1\rangle&\equiv&
\hat{T}|{\rm VB}_1\rangle+\frac{N}{2}|{\rm VB}_1\rangle
=\sum_{l={\rm even}}\hat{t}_{l,l+1}|{\rm VB}_1\rangle\\
|\phi_2\rangle&\equiv&
\hat{T}|{\rm VB}_2\rangle-\frac{N}{2}|{\rm VB}_2\rangle
=-\sum_{l={\rm odd}}\hat{t}_{l,l+1}|{\rm VB}_2\rangle.
\end{eqnarray*}
Then,
\begin{eqnarray*}
\langle{\rm VB}_i|\phi_j\rangle\simeq\delta_{i,j}(-1)^{i+1}\frac{N}{8},
\end{eqnarray*}
which gives
\begin{eqnarray*}
\langle{\rm VB}_i|\hat{T}|{\rm VB}_j\rangle\simeq\delta_{i,j}(-1)^i\frac{3N}{8}.
\end{eqnarray*}
Therefore,
\begin{eqnarray*}
\langle \Psi|\hat{T}|\Psi\rangle\simeq0.
\end{eqnarray*}

Also, we can show
\begin{eqnarray*}
\langle\phi_i|\phi_j\rangle\simeq\delta_{i,j}
\Big(\frac{N^2}{64}+\frac{3N}{32}\Big),
\end{eqnarray*}
which gives
\begin{eqnarray*}
\langle{\rm VB}_i|\hat{T}^2|{\rm VB}_j\rangle\simeq
\delta_{i,j}
\Big(
\frac{9N^2}{64}
+\frac{3N}{32}
\Big).
\end{eqnarray*}
Therefore,
\begin{eqnarray*}
\langle \Psi|\hat{T}^2|\Psi\rangle=O(N^2).
\end{eqnarray*}
Hence we have shown Eq.~(\ref{rfluc}).~\rule[-2pt]{5pt}{10pt}


\end{document}